\documentstyle[12pt,epsf]{article}
\newcommand{\beq}{\begin{equation}}
\newcommand{\eeq}{\end{equation}}
\newcommand{\beqa}{\begin{eqnarray}}
\newcommand{\eeqa}{\end{eqnarray}}
\newcommand{\ba}{\begin{array}}
\newcommand{\ea}{\end{array}}
\newcommand{\CR}{\nonumber \\}
\newcommand{\pa}{\partial}
\newcommand{\A}{\alpha}
\newcommand{\B}{\beta}
\newcommand{\D}{\delta}

\newcommand{\La}{\Lambda}
\newcommand{\p}{\Phi}

\newcommand{\lm}{\lambda}

\newcommand{\half}{{1\over 2}}
\newcommand{\Tr}{{\rm Tr}}

\newcommand{\cdm}{{\cal D}_{\mu}}
\newcommand{\ps}{\psi}
\newcommand{\bs}{\bar{\sigma}}
\newcommand{\eq}{\begin{equation}}
\newcommand{\en}{\end{equation}}
\newcommand{\eqn}{\begin{eqnarray}}
\newcommand{\enn}{\end{eqnarray}}
\newcommand{\th}{\theta}
\newcommand{\thb}{\bar{\theta}}
\newcommand{\sg}{\sigma}
\newcommand{\LM}{\Lambda}
\newcommand{\da}{\dagger}

\begin{document}

\makeatletter
\def\setcaption#1{\def\@captype{#1}}
\makeatother

\begin{titlepage}
\null
\begin{flushright} 
hep-th/0002119  \\
UT-876  \\
February, 2000 
\end{flushright}
\vspace{0.5cm} 
\begin{center}
%{\Large \bf
{\LARGE %\bf
A Note on Superfields and 
Noncommutative Geometry
\par}
\lineskip .75em
\vskip2.5cm
\normalsize
{\large Seiji Terashima}\footnote{
E-mail:\ \ seiji@hep-th.phys.s.u-tokyo.ac.jp} 
\vskip 1.5em
{\large \it  Department of Physics, Faculty of Science, University of Tokyo\\
Tokyo 113-0033, Japan}
\vskip3cm
{\bf Abstract}
\end{center} \par
We consider the supersymmetric field theories
on the noncommutative ${\bf R}^4$ using the superspace formalism
on the commutative space.
The terms depending on the parameter of the noncommutativity $\Theta$
are regarded as the interactions.
In this way we construct the $N=1$ supersymmetric action 
for the $U(N)$ vector multiplets and chiral multiplets 
of the fundamental, anti-fundamental and adjoint representations
of the gauge group.
The action
for vector multiplets 
of the products gauge group and its bi-fundamental matters
is also obtained.
%We find that the scalar potentials are characteristic forms 
%for the D and F terms and
We discuss the problem of the derivative terms of the auxiliary fields.

\end{titlepage}

\baselineskip=0.7cm

%%%%%%%%%%%%%%%%%%%%%%%%%%%%%%%%%%%%%%%%%%%%%%%%%%%%
%\section{Introduction}
%%%%%%%%%%%%%%%%%%%%%%%%%%%%%%%%%%%%%%%%%%%%%%%%%%%%

In the past few years there has been much development in our understanding
of the dynamics of supersymmetric gauge theories and
superstring theories.
Among these 
it has been discovered that 
the noncommutative gauge theories
naturally appear \cite{CoDoSc,DoHu} 
when the D-branes with constant $B$ fields is considered.

Recently Seiberg and Witten have argued that
the noncommutative gauge theories realized as effective theories 
on D-branes are equivalent to
some ordinary gauge theories \cite{SeWi}.
In a single D-brane case, they have shown that 
the effective action for the D-brane is consistent 
with the equivalence if all derivative terms are neglected.
Furthermore, it has been shown that 
the D-brane action, including derivative terms, 
computed in the string theory
is consistent with the equivalence
if we keep the two derivative terms but
neglect the fourth and higher order derivative terms
\cite{Te2,OkTe,Ok}.

For deeper understanding for these phenomena,
it is natural to investigate the field theories
on the noncommutative geometry by
the field theoretical approaches.
In particular by the perturbative analysis
it was found in \cite{MiSeVa} 
that 
the IR effects and UV effects are mixed in the noncommutative field
theory.
%\footnote{
%The perturbative aspects of noncommutative field theories
%are discussed in many people \cite{Fi} }

To proceed further
it may be important to study the noncommutative field theories 
with supersymmetry
since their actions are highly constrained and
we may understand the dynamics of these theories.
To obtain the supersymmetric action, 
superfields on the noncommutative geometry may be desired.
Explicit two-dimensional $N=1$ noncommutative superspace was obtained in
\cite{FaZa}.

However, in this note,
instead of investigating the noncommutative superspace formalism,
we consider the ordinary superspace and superfields \cite{SaSt}
and represent the noncommutative field theory using these notions.
From the commutative supersymmetric action
written by the superfields,
we can obtain the noncommutative supersymmetric action 
by replacing the ordinary product between superfields
to the $*$ product 
defined by the formula
\beq
\left.
f(x)*g(x)=e^{\frac{i}{2} \Theta^{ij}
\frac{\pa}{\pa \xi^i} \frac{\pa}{\pa \zeta^j} } f(x+\xi) g(x+\zeta )
\right|_{\xi=\zeta=0}.
\eeq
Here we regard the additional terms depend on $\Theta^{ij}$
as the interaction terms with derivatives
although we do not expand the $*$ product 
explicitly.
We can do so because
if $\p$ is superfield then $\pa^n \p$ is also superfields.
Thus it is obvious that this action has the supersymmetry 
and the R symmetry when it exists in the action with $\Theta=0$
because $\pa_\mu \theta=0$ where $\theta$ is  the fermionic coordinate.
This treatment of the superfields
is similar to the notions of superspace and
superfields in noncommutative geometry \cite{ChZa}, in which 
only the chiral superfields has been considered.

In this paper, 
we consider the $N=1$ supersymmetric theories on the
noncommutative ${\bf R}^4$ 
although the above observation does not depend on the dimension of 
the spacetime and the number of the supersymmetries.
We construct the $N=1$ supersymmetric action 
for the $U(N)$ vector multiplets and chiral multiplets 
of the fundamental, anti-fundamental and adjoint representations
of the gauge group.
The actions
for gauge fields of the products gauge groups and its bi-fundamental
matters are also obtained.

It is argued that 
even if we do not require the supersymmetry,
only these gauge groups and the matters
are possible for the noncommutative gauge theories.
We also find that the scalar potentials have some characteristic forms 
and
discuss the problem of the derivative terms of the auxiliary fields.

The convention and notation taken in this paper 
are same as in \cite{WeBa}
except for the spacetime indices, which are denote by $\mu, \nu, \cdots$ 
in this paper, and the gauge field $A_\mu$.

First we consider the chiral superfields which satisfy 
$\bar{D}_{\dot{\A}} \Phi =0$.
Using the coordinate
$y^{m}=x^{m}+i \th \sg^{m} \thb $,
these are written as 
$\Phi(y,\th,\thb) = A(y)+\sqrt{2} \th \ps(y)+ \th\th F(y)$.
The supersymmetry transformations are identical for commutative
counterparts 
\eqn
\D_\xi A & = & \sqrt{2} \xi \ps, \CR
\D_\xi \ps & = & i \sqrt{2} \sg^m \bar{\xi} \pa_m A+\sqrt{2} \xi F, \CR
\D_\xi F & = & i \sqrt{2} \bar{\xi} \bar{\sg}^m \pa_m \ps,
\label{st}
\enn
because we simply consider the ordinary superfields.

Defining 
\beq
(\prod_{i=1}^n f_i)_*=f_1 * f_2 * \cdots * f_n,
\eeq
the most generic action which can be constructed from 
the chiral superfields ${\Phi^i}$ takes the form
\eq
S = \int d^4 x \left(
\int d^2 \th d^2 \thb K(\Phi^i,\Phi^{ \dagger j})_*
      + \left[\int d^2 \th W(\Phi^i)_* +h.c. \right]
\right),
\en
where $  \int d^2 \th \, \th^2=1$ and 
$  \int d^2 \thb \, \thb^2=1$.
This is invariant under
$K(\Phi^i,\Phi^{ \dagger j})_* \rightarrow 
K(\Phi^i,\Phi^{ \dagger j})_*+F(\p)_*+F(\p^\da)_*^\da$.
Although the action of the component fields can be obtained 
straightforwardly,
we will only give some examples below.

%\eqn
%\cL & = & -g_{i j^*} \pa_m A^i \pa^m A^{*j} - i g_{i j^*} 
%\bar{\chi}^j \bar{\sg}^m D_m \chi^i \CR
%& & +\frac{1}{4} R_{i j^* k l^*}  \chi^i \chi^k \bar{\chi}^j \bar{\chi}^l \CR
%& & -\frac{1}{2} D_i D_j W \chi^i \chi^j -
%\frac{1}{2} D_{i^*} D_{j^*} W^* \bar{\chi}^i \bar{\chi}^j \CR
%& & -g^{i j^*} D_i W D_{j^*} W^*
%\enn
%\eqn
%D_i W & = & \frac{\pa}{\pa A^i} W \CR
%D_i D_j W & = & \frac{\pa^2}{\pa A^i \pa A^j} W - 
%\Gamma^k_{ij} \frac{\pa}{\pa A^k}W
%\enn
%\eqn
%g_{i j^*} & = & \frac{\pa}{\pa A^i} \frac{\pa}{\pa A^{* j}} K(A) \CR
%g_{i j^*,k} & = & \frac{\pa}{\pa A^k} g_{i j^*}=g_{m j^*} \Gamma^m_{i k} \CR
%g_{i j^*,k*} & = & 
%\frac{\pa}{\pa A^{* k}} g_{i j^*}=g_{i m^*} \Gamma^{m^*}_{i^*  k^*} 
%\enn

First we consider 
the action with $K=\p^\da *\p +a \p*\p*(\p^\da)+a^\da
\p*(\p^\da)*(\p^\da)$ and $W=0$,
where $a$ is some numerical constant.
Note that 
\beq
(A*B)^\da =B^\da * A^\da.
\eeq
The part of the action which depends on $F$ becomes 
\beqa
S |_{F} &=& \int d^4 x \left( F^\da F+(a A*F*F^\da+a F*A*F^\da+a F*F*A^\da +h.c)
\right) \CR
&=& \int d^4 x \left( F^\da F+(a F (F^\da*A) +a F (A*F^\da)+a F (F*A^\da) +h.c)
\right).
\eeqa
Here we have used that
\beq
\int d^4 x A*B=\int d^4 AB =\int d^4 x B*A,
\eeq
which implies that the integral of the product of fields 
with $*$ product 
are unchanged by the cyclic rotation of the fields.
Thus the action clearly contains
the derivative of the auxiliary field $F$ and
it is difficult to eliminate it from the action 
using the equation of motion.
Moreover $F$ may become the propagating field if
the noncommutative parameter $\Theta^{0 \mu} \neq 0$ for some $\mu$.
To avoid these problems,
we only consider the canonical K\"{a}hler potential
$K=\sum_i \p_i^\da * \p_i$
below.

With this $K$, the action with non vanishing superpotential
does not have the derivative of $F$ then $F$ can be eliminated.
This can be seen from the fact that the terms which depend on $F$ 
in the superpotential are linear in $F$.
For example, 
the $F$ dependent parts of the action with $W=a \p^n$ become
\beq
\int d^4 x \left( \half F^\da F+a \sum_{i=1}^n 
F \,\, (A^{n-i})_* * (A^{i-1})_*+h.c) \right).
\eeq

Next we consider the noncommutative Wess-Zumino model 
\cite{WeZu} \cite{ChRo} \cite{Go}
\beq
S_{WZ} = \! \int d^4 x \left(
\int d^2 \th d^2 \thb \, \p_i^\da \!*\! \p_i
      + \left[\int d^2  \th \left(  \half m_{ij} \p_i \!*\! \p_j 
+\frac{1}{3} g_{ijk} 
\p_i \!*\! \p_j \!*\! \p_k +g_i \p_i \right)   +h.c. \right]
\right),
\eeq
where the mass $m_{ij}$ is symmetric in their indices, however,
the coupling $g_{ijk}$ is not necessarily symmetric.
By trancing the procedure for the $\Theta=0$ case,
we can easily find that 
\beqa
S_{WZ} \!\!\!\! &=& \!\!\!  \int d^4 x \left(
- \pa_\mu A_i^\da \pa^\mu A_i +i \pa_\mu \ps_i^\da \bs^\mu \ps_i
+F^\da_i F_i  \right)
\CR
&& 
\!\!\!\!\!\!\!\!  
+ \!\! \int d^4 x 
\left[ 
\frac{1}{3} g_{ijk} 
\left(
F_i \,  A_j \!*\! A_k + F_j \,  A_k  \!*\! A_i +F_k \, A_i \!*\! A_j    
\!\! -A_i \, \ps_j \!*\! \ps_k 
\!-\! A_j \,  \ps_k  \!*\! \ps_i \!-\! A_k \, \ps_i \!*\! \ps_j    
\right)
\right. \CR
&& \left. 
+g_i F_i 
+m_{ij} \left( A_i F_j -\half \ps_i \ps_j \right) 
+h.c \right]. 
\label{WZ}
\eeqa
The equation of motions of $F_i$ is 
\beq
F_i^\da=g_i+m_{ij} A_j +\frac{1}{3}\left( g_{ijk} 
+ g_{kij}   +g_{jki} \right) A_j \!*\! A_k ,
\eeq
and the supersymmetry transformation becomes (\ref{st}) with this $F_i$.
We note that the typical scalar potential has 
the form $A^\da \!*\!  A^\da \!*\!  A \!*\! A$
and the notion of holomorphy is still valid at $\Theta \neq0$.

Now we consider the vector superfields $V=V^\da$ \cite{WeZu2,FeZu} , 
\eqn
\label{vnotenkai}
V(x,\th,\thb) & = & C(x)+i\th \chi(x)-i\thb \bar{\chi} (x) \CR
               & & +\frac{i}{2} \th \th \left[ M(x)+iN(x)\right] -
               \frac{i}{2} \thb \thb \left[ M(x)-iN(x)\right]  \CR
               & & -\th \sg^\mu \thb A_\mu (x) 
               +i \th \th \thb \left[ \bar{\lm}(x)+\frac{i}{2} 
\bar{\sg}^\mu \pa_\mu \chi(x) \right] \CR
              & & - i \thb \thb \th 
\left[ \lm (x)+ \frac{i}{2} \sg^\mu \pa_\mu \bar{\chi} (x) \right] 
              + \frac{1}{2} \th \th \thb \thb 
\left[ D(x) + \frac{1}{2} \Box C(x) \right].
\enn   
Since the $*$ product contain derivative,
the ordinary gauge invariant action for the vector superfields
can not be generalized 
to $\Theta \neq 0$ case.
Then we should introduce noncommutative gauge field
$A_\mu=T^a A^a_\mu$, where $T^a$ is the matrix 
for a representation of the gauge group $G$ and satisfies that
$(T^a)^\da=T^a$ and $\Tr (T^a T^b)=k$.

Hereafter we briefly discuss the some properties of the noncommutative 
gauge field without requiring supersymmetry in order to prepare to treat 
the vector superfield.
We assume that
the noncommutative gauge transformation is the naive generalization of 
the ordinary Non-Abelian gauge transformation,
\beq
A_\mu \rightarrow U \!*\! A_\mu \!*\! U^{-1} + i U \!*\! \pa_\mu U^{-1},
\eeq
where $U=(e^{i \lm})_*$, $U^{-1}=(e^{-i \lm})_*=U^\da$ and
$\lm =T^a \lm^a=\lm^\da$.
The infinitesimal version of this is 
\beqa
\D_\lm A_\mu \!\!
&=& \pa_\mu \lm + i \lm \!*\! A_\mu- i A_\mu \!*\! \lm \CR
&=& T^a (\pa_\mu \lm^a) 
+\frac{i}{2} [ T^a, T^b ] (\lm^a \!*\! A_\mu^b + A_\mu^b\!*\! \lm^a )
+\frac{i}{2} \{ T^a, T^b \} (\lm^a \!*\! A_\mu^b - A_\mu^b\!*\! \lm^a ).
\label{igt}
\eeqa
From this if $\{ T^a, T^b \}$ is not a linear combination of 
$T^d$ for some $a,b$, the gauge transformation is not closed.
Thus the noncommutative gauge transformation is consistent
only for unitary group $G=U(N)$ or its direct product 
$G=\prod_a U(N_a)^{(a)}$.
In addition to this restriction,
we should take $T^a$ as the matrix for 
the fundamental or anti-fundamental representation
by the requirement of the closure of (\ref{igt}).
However $T^a$ and $\tilde{T}^a=-{}^t T$, 
which represent the fundamental and
anti-fundamental representations respectively,
give the different gauge transformations via (\ref{igt}).
Then we take 
$(T^a)^{i}_{\,\, j}$ to be the matrix for the fundamental representation,
$N \times N$ Hermitian matrix.\footnote{
Of course we can choose the matrix of the anti-fundamental representation
instead of the one for fundamental representation.}

We can see
that
the representation of the gauge group $G$ of the matter is also 
restricted to fundamental ($\bf N$), anti-fundamental ($\bar{\bf N}$), 
adjoint (${\bf N} \times {\bar{\bf N}}$) 
or bi-fundamental (${\bf N} \times {\bar{\bf M}}$)
from the consideration of the possible form of 
the covariant derivative. In \cite{Ha} this restriction has been shown
for $G=U(1)$ case, where $T^1=k^{\half}$.
For the fundamental matter represented as the column vector 
$(\ps)^i=\ps^i$,
the gauge transformation 
and the covariant derivative
are given by $\ps \rightarrow U \!*\! \ps$ and
\beq
\cdm \ps = \pa_\mu \ps-i A_\mu \!*\! \ps,
\eeq
respectively.
We can easily check $\cdm \ps \rightarrow U \!*\! (\cdm \ps)$
under the gauge transformation.
Noting $\cdm \ps^\da \equiv (\cdm \ps)^\da 
= \pa_\mu \ps+i \ps^\da \!*\!  A_\mu$,
the covariant derivative for the anti-fundamental matter 
$(\tilde{\ps})_i=\tilde{\ps}_i$, which is transformed as
$ \tilde{\ps} \rightarrow \tilde{\ps} \!*\! U^{-1}$,
is 
\beq
\cdm \tilde{\ps}=\pa_\mu \tilde{\ps}+i \tilde{\ps} \!*\!  A_\mu.
\eeq
The adjoint matter $(\ps_{adj})_{\,\, i}^{j}$, which is transformed as
$ \ps_{adj}  \rightarrow  U \!*\! \ps_{adj} \!*\! U^{-1}$, can have the
covariant derivative
\beq
\cdm \ps_{adj} = \pa_\mu \ps_{adj}-i A_\mu \!*\! \ps_{adj}
+i \ps_{adj} \!*\!  A_\mu.
\eeq
This is also seen from the covariant derivative 
for the bi-fundamental matter $(\ps_{N\bar{M}})_{\,\, i}^{j}$,
where $j=1 \ldots N$ and $i=1 \ldots M$,
is
\beq
\cdm \ps_{N\bar{M}} = \pa_\mu \ps_{N\bar{M}}-i A_\mu^{(1)} 
\!*\! \ps_{N\bar{M}} +i \ps_{N\bar{M}} \!*\!  A_\mu^{(2)}.
\eeq
Here $A_\mu^{(1)} $ and $A_\mu^{(2)} $ are the gauge fields for 
$U(N)$ and $U(M)$ respectively and
the gauge transformation for it is given by 
$ \ps_{N\bar{M}}  \rightarrow  U^{(1)} \!*\! \ps_{N\bar{M}} \!*\!
{U^{(2)}}^{-1}$.

The interaction terms are severely constrained by the gauge symmetry
and the possible forms of the terms are the
polynomials of $\tilde{\ps} \!*\! ({\ps_{adj}}^n)_* \!*\! \ps$, and 
$\Tr ({\ps_{adj}}^n)_*$
for $G=U(N)$.
For the product group case,
there are other terms which are allowed by the symmetry.

On the basis of this observation, 
we return to consider the vector superfields $V=T^a V^a$.
We define the noncommutative super gauge transformation as
\beq
(e^{-2V})_* \rightarrow (e^{-2V'})_* =
(e^{-i \La^\da})_* * (e^{-2V})_* * (e^{i \La})_*.
\eeq
The chiral superfield 
\eq 
W_\A=  \frac{1}{8} \bar{D} \bar{D} 
\left( (e^{2 V})_* * D_\A (e^{-2 V})_* \right),
\en
is transformed as
\eq
W_\A \rightarrow 
{W_\A}^\prime=(e^{-i \LM})_* * W_\A * (e^{i \LM})_*.
\en

Because of 
\eq
V^\prime=V+i(\LM-\LM^ \dagger )+\cdots,
\en
we can choose the Wess-Zumino gauge 
in which $C,\chi,M,N$ are eliminated.
In the Wess-Zumino gauge we see 
\beq
W_\A(y)=-i \lm_\A (y) + \th_\A D(y)
-\frac{i}{2} (\sg^\mu \bs^\nu \th)_\A F_{\mu \nu} (y) 
+ \th^2 (\sg^\mu \cdm \bar{\lm} (y))_\A,
\eeq
where 
\beq
F_{\mu \nu}=\pa_\mu A_\nu -\pa_\nu A_\mu-i A_\mu * A_\nu +i A_\nu * A_\mu,
\eeq
and
\beq
\cdm \bar{\lm} 
=\pa_\mu \bar{\lm} -i A_\mu * \bar{\lm}+i \bar{\lm} * A_\mu.
\eeq

Defining the complex gauge coupling 
$\tau=\frac{\tilde{\th}}{2 \pi}+\frac{4 \pi i}{g^2}$,
we obtain the action of 
the noncommutative supersymmetric $U(N)$ gauge field theory,\footnote{
Although it is straightforward to obtain the action
for the case of general K\"{a}hler  potential,
we only consider here the canonical K\"{a}hler  potential 
in order to avoid
the problem with the derivative terms of the auxiliary fields.}
\beqa
S_V &=& \frac{1}{16 \pi k} \int d^4 x d \th^2 \Tr \left(
-i \tau W^\A * W_\A +h.c. \right) \CR
&=& \int d^4 x \Tr \left(
-\frac{1}{4 g^2} F^{\mu \nu} F_{\mu \nu} 
-\frac{i}{g^2} \lm \sg^\mu (\cdm \bar{\lm} )+
\frac{1}{2 g^2} D D -\frac{\tilde{\th}}{64 \pi^2} 
\epsilon_{\mu \nu \A \B} F^{\mu \nu} F^{\A \B}
\right).
\eeqa
Note that the definitions of $F_{\mu \nu}$ and $\cdm$
depend on $\Theta$.

Next we consider the chiral superfields coupled to the vector superfields.
The gauge transformations for the fundamental, anti-fundamental,
adjoint and bi-fundamental chiral superfields are given by 
\beqa
\p & \rightarrow & (e^{-i \La})_* * \p, \CR
\tilde{\p} & \rightarrow &  \tilde{\p} * (e^{i \La})_*, \CR
\p_{adj} & \rightarrow & 
(e^{-i \La})_* * \p_{adj} * (e^{i \La})_* , \CR
\p_{N\bar{M}} & \rightarrow & 
(e^{-i \La^{(1)}})_* * \p_{N\bar{M}} * (e^{i \La^{(2)}})_* ,
\eeqa
respectively.

We can see that 
the supersymmetric gauge invariant actions including kinetic terms
are
\beqa
S_{\p} &=& \int d^4 x d^2 \th d^2 \thb 
\left( \p^\da * (e^{-2 V})_* * \p \right) \CR
&=& \int d^4 x \left( - (\cdm A^\da) ({\cal D}^{\mu} A) 
-i \ps^\da \bs^\mu (\cdm \ps)-A^\da *D*A  \right. \CR
&& \left.  \hspace{1cm}
-i \sqrt{2} A^\da * \lm * \ps 
+ i \sqrt{2} \ps^\da * \lm^\da * A+F^\da F
\right), \CR
S_{\tilde{\p}} &=& \int d^4 x d^2 \th d^2 \thb 
\left( \tilde{\p} * (e^{2 V})_* * \tilde{\p}^\da \right) \CR
&=& \int d^4 x \left( - (\cdm \tilde{A}) ({\cal D}^{\mu} \tilde{A}^\da) 
-i \tilde{\ps} \sg^\mu (\cdm \tilde{\ps}^\da )+\tilde{A} *D*\tilde{A}^\da
\right. \CR
&& \left.  \hspace{1cm}
-i \sqrt{2} \tilde{A} * \lm^\da * \tilde{\ps}^\da 
+ i \sqrt{2} \tilde{\ps} * \lm * \tilde{A}^\da
+\tilde{F} \tilde{F}^\da \right), \CR
S_{\p_{adj}} &=& \int d^4 x d^2 \th d^2 \thb \frac{1}{k} \Tr 
\left( (e^{2 V})_* * \p_{adj}^\da * (e^{-2 V})_* * \p_{adj} \right) \CR
&=& \int d^4 x \frac{1}{k} \Tr 
\left( - (\cdm A_{adj}^\da) ({\cal D}^{\mu} A_{adj}) 
-i \ps_{adj}^\da \bs^\mu (\cdm \ps_{adj})
\right. \CR
&& \left. \hspace{1cm}
-A_{adj}^\da \!*\! D \!*\! A_{adj}
+A_{adj} \!*\! D\!*\! A_{adj}^\da  \right. \CR
&& \left. \hspace{1cm} 
-i \sqrt{2} A_{adj}^\da \!*\!  \lm \!*\!  \ps_{adj} 
+ i \sqrt{2} \ps_{adj}^\da \!*\!  \lm^\da \!*\!  A_{adj}
\right. \CR
&& \left. \hspace{1cm}
-i \sqrt{2} A_{adj} \!*\!  \lm^\da \!*\!  \ps_{adj}^\da 
+ i \sqrt{2} \ps_{adj} \!*\!  \lm \!*\!  A_{adj}^\da 
+F_{adj}^\da F_{adj}
\right), \CR
S_{\p_{N\bar{M}}} &=& \int d^4 x d^2 \th d^2 \thb \Tr 
\left( 
(e^{2 V^{(2)}})_* * \p_{N\bar{M}}^\da * (e^{-2 V^{(1)}})_* 
* \p_{N\bar{M}} \right) \CR
&=& \int d^4 x \Tr 
\left( - (\cdm A_{N\bar{M}}^\da) ({\cal D}^{\mu} A_{N\bar{M}}) 
-i \ps_{N\bar{M}}^\da \bs^\mu (\cdm \ps_{N\bar{M}})
\right. \CR
&& \left. \hspace{1cm}
-A_{N\bar{M}}^\da \!*\! D^{(1)} \!*\! A_{N\bar{M}}
+A_{N\bar{M}} \!*\! D^{(2)} \!*\! A_{N\bar{M}}^\da  \right. \CR
&& \left. \hspace{1cm} 
-i \sqrt{2} A_{N\bar{M}}^\da \!*\!  \lm^{(1)} \!*\!  \ps_{N\bar{M}} 
+ i \sqrt{2} \ps_{N\bar{M}}^\da \!*\!  (\lm^{(1)})^\da \!*\!  A_{N\bar{M}}
\right. \CR
&& \left. \hspace{1cm}
-i \sqrt{2} A_{N\bar{M}} \!*\!  (\lm^{(2)})^\da \!*\!  \ps_{N\bar{M}}^\da 
+ i \sqrt{2} \ps_{N\bar{M}} \!*\!  \lm^{(2)} \!*\!  A_{N\bar{M}}^\da 
+F_{N\bar{M}}^\da F_{N\bar{M}}
\right).
\label{S4}
\eeqa

In the $\Theta=0$ case
$\Tr(e^{2V} \p_{adj}^\da e^{-2V} \p_{adj})/k$ is equivalent 
to 
\beq
\sum_{a,b=1}^{N^2} {\p_{adj}^a}^\da 
e^{-2 \sum_{c=1}^{N^2} V^c (T^c_{adj})_{ab}} \p_{adj}^b,
\label{la}
\eeq
where $T^c_{adj}$ is the matrix of the adjoint representation and
we have used $ e^Y X e^{-Y} =X+[X,Y]+\half [Y [Y,X]] + \cdots$.
However 
the generalization of (\ref{la}) to the $\Theta \neq 0$ case
is not noncommutative gauge invariant. 
For the anti-fundamental chiral superfield
the similar phenomena can be shown and
in general we should use the matrix of 
the fundamental representation $T^a$ only.

We note that there are no derivative terms
of the auxiliary fields $D$ in the actions (\ref{S4})
and typical scalar potential are the form of $A^\da A A^\da A$,
which is different from the one from the superpotential.
In \cite{ArBeKo} it has been shown 
that the noncommutative complex scalar field theory
with the  interaction $A^\da A A^\da A$ does not
suffer from IR divergences at one-loop insertions level.
It is also seen that 
the classical moduli space of vacua is unchanged by varying $\Theta$.

Finally as in the commutative case 
we can obtain the transformation of the supersymmetry
in the Wess-Zumino gauge 
\eqn
\D_{\xi} A & = & \sqrt{2} \xi \ps, \CR
\D_{\xi} \ps & = & i \sqrt{2} \sg^\mu \bar{\xi} 
(\cdm A) + \sqrt{2} \xi F, \CR
\D_{\xi} F & = & i \sqrt{2} \bar{\xi} \bar{\sg}^\mu (\cdm \ps)
-2 i \bar{\xi} \bar{\lm} A,  \CR
\D_{\xi} A_\mu & = & -i \bar{\lm}  \bar{\sg}^\mu \xi +
i \bar{\xi} \bar{\sg}^\mu  \lm, \CR
\D_{\xi} \lm & = & \sg^{\mu \nu} \xi F_{\mu \nu} + i \xi D, \CR
\D_{\xi}  D & = & - \xi \sg^\mu (\cdm \bar{\lm})
- (\cdm \lm^{(a)}) \sg^\mu \bar{\xi}.
\enn
This formula is valid for 
the chiral superfields of any representation of $G$.

The form of the noncommutative gauge invariant superpotential 
are constrained
as stated for the component fields.
Using (\ref{WZ}),
we can easily write down the action 
of the component fields
for any superpotential which is renormalizable at $\Theta=0$.

It is possible to generalize these considerations to
the extended superspace.
On the other hand, we can construct the action with
the extended supersymmetry by the $N=1$ superfields for commutative
case.
We can easily construct the action with non-vanishing $\Theta$
corresponding to the commutative action with
the extended supersymmetry.
Even at $\Theta \neq 0$, these action has the extended supersymmetry
because of the existence of the 
R symmetry which rotates the generators of the supersymmetry.
In fact the $U(N)$ noncommutative gauge theory with
one adjoint chiral superfield, $N_f$ fundamental 
and $N_f$ anti-fundamental
chiral superfields and $W=\sqrt{2} \sum_{i=1}^{N_f} 
\tilde{\p}_{(i)} * \p_{adj} * \p_{(i)}$ has $N=2$ supersymmetry.
We can also obtain the noncommutative $N=4$ supersymmetric action
with $W= \Tr ( \p_{adj}^{(1)} * \, 
(\p_{adj}^{(2)} * \p_{adj}^{(3)} -\p_{adj}^{(3)} * \p_{adj}^{(2)}) 
\, )$, where $\p_{adj}^{(i)}$ are three adjoint chiral superfields.

The effective theories of the D-branes on the orbifold 
are the quiver gauge theories \cite{DoMo}
or the elliptic models \cite{Wi} which have bi-fundamental matter.
Thus it is interesting that the action for 
the supersymmetric gauge theories with bi-fundamental matters
can be constructed.

%%%%%%%%%%%%%%%%%%%%%%%%%%%%%%%%%%%%%%%%%%%%%%%%%%%%
%\section{Conclusion}
%%%%%%%%%%%%%%%%%%%%%%%%%%%%%%%%%%%%%%%%%%%%%%%%%%%%

In this paper, 
we have considered the $N=1$ supersymmetric theories on the
noncommutative ${\bf R}^4$.
We have constructed the $N=1$ supersymmetric action 
for the $U(N)$ vector multiplets and chiral multiplets 
of the fundamental, anti-fundamental and adjoint representations
of the gauge group.
The actions
for gauge fields of the products gauge groups and its bi-fundamental
matters have also been obtained.
We have been argued that 
only these gauge groups and the matters
are possible for the noncommutative gauge theories.
We have also found that the scalar potentials 
have some characteristic forms 
and
discussed the problem of the derivative terms of the auxiliary fields.

It is interesting to generalize
the results obtained in this paper to
the nonlinearly realized supersymmetry.
This is important because the supersymmetric DBI action
which is the effective theory on a D-brane has this symmetry \cite{BaGa}
and has been used for the instanton in the D-brane with 
the $B$ field \cite{SeWi,Te} which is related to 
the noncommutative instanton \cite{NeSh}.

\vskip6mm\noindent
{\bf Acknowledgements}

\vskip2mm
I would like to thank T. Kawano, T. Yanagida and S-K. Yang
for useful conversations.
I would also like to thank K. Kurosawa and Y. Nomura 
for discussions.
This work was supported in part by JSPS Research Fellowships for Young 
Scientists. \\

\noindent
{\bf Note added}: 

As this article was being completed,
we received the preprint
\cite{FeLl} which substantially overlap the present work.

\newpage

%%%%%%%  References

\end{document}